\documentstyle[12pt,a4wide]{article}
\input epsf

\newcommand{\news}{\setcounter{equation}{0}}
\newcommand{\be}{\begin{equation}}
\newcommand{\ee}{\end{equation}}
\newcommand{\bea}{\begin{eqnarray}}
\newcommand{\eea}{\end{eqnarray}}
\newcommand{\bean}{\begin{eqnarray*}}

\newcommand{\eean}{\end{eqnarray*}}
\newcommand{\beq}{\begin{equation}}
\newcommand{\eeq}{\end{equation}}
\font\upright=cmu10 scaled\magstep1
\font\sans=cmss12
\newcommand{\ssf}{\sans}
\newcommand{\stroke}{\vrule height8pt width0.4pt depth-0.1pt}
\newcommand{\Z}{\hbox{\upright\rlap{\ssf Z}\kern 2.7pt {\ssf Z}}}

\newcommand{\C}{{\rlap{\rlap{C}\kern 3.8pt\stroke}\phantom{C}}}
\newcommand{\R}{\hbox{\upright\rlap{I}\kern 1.7pt R}}
\newcommand{\CP}{\C{\upright\rlap{I}\kern 1.7pt P}}
\newcommand{\mmm}{(2-m)(2m-1)(m+1)}

{\hfill $\square$ \end{trivlist}}

\begin{document}
\pagestyle{plain}
\title{\vskip -70pt
\begin{flushright}
{\normalsize MS-96-016} \\
{\normalsize UKC/IMS/96-64}
\end{flushright}
\vskip 20pt
{\bf \Large \bf A Monopole Metric}
 \vskip 10pt
}
\author{H. W. Braden$^{\ \dagger}$
and 
P. M. Sutcliffe$^{\ \ddagger}$
\thanks{This work was supported in part by the 
Nuffield  Foundation}\\[10pt]
{\normalsize
$\dagger$ {\sl Department of Mathematics and Statistics,}}\\
{\normalsize {\sl The University of Edinburgh, Edinburgh, UK. }}\\
{\normalsize {\sl Email hwb@ed.ac.uk}}\\[10pt]
{\normalsize 
$\ddagger$ {\sl
Institute of Mathematics,
University of Kent at Canterbury,}}\\
{\normalsize {\sl Canterbury CT2 7NZ, England.}}\\
{\normalsize {\sl Email P.M.Sutcliffe@ukc.ac.uk}}\\[10pt]
}

\date{October 1996}
\maketitle

\begin{abstract}
\noindent We calculate explicitly in terms of complete elliptic integrals
the metric on the moduli space of tetrahedrally-symmetric, charge four,
$SU(2)$ monopoles. Using this we verify that in the asymptotic regime
 the metric of Gibbons and Manton is exact up to
exponentially suppressed corrections.
\end{abstract}
\newpage
\renewcommand{\thepage}{\arabic{page}}

\section{Introduction}
\news
The moduli space ${\cal M}_n$ of charge $n$ $SU(2)$ BPS monopoles is
a $4n$-dimensional manifold, whose metric is of interest
for three main reasons. First, it is known to be hyperk\"ahler,
and explicit examples of such metrics are rare. Second, the dynamics
of  $n$ slowly moving monopoles can be approximated by geodesic motion
on  ${\cal M}_n$ \cite{M,St}. Finally, following the work of Sen \cite{Sen},
predictions of S-duality in the $N=4$ supersymmetric quantum theory
can be tested by an analysis of the harmonic forms on ${\cal M}_n$.

There is an isometric splitting
\be
 \widetilde{\cal M}_n=\R^3\times\mbox{S}^1\times {\cal M}_n^0
\ee
where $\widetilde{{\cal M}}_n$ is an $n$-fold covering of ${\cal M}_n$.
Thus the interesting structure of the moduli space is contained in the
$4(n-1)$-dimensional hyperk\"ahler manifold ${\cal M}_n^0.$ 
The simplest case ${\cal M}_2^0$ is the Atiyah-Hitchin manifold \cite{AH},
where the metric can be written explicitly in terms of complete elliptic
integrals. Unfortunately, for $n>2$ the metric on ${\cal M}_n^0$ is not known.
Indeed until recently the exact metric was unknown on any submanifold
of ${\cal M}_n^0$. However, recent results \cite{HSd,B} have shown that
for $n>2$, ${\cal M}_n^0$ contains a totally geodesic submanifold which
is the Atiyah-Hitchin manifold. Thus, in this sense, the metric on a
4-dimensional submanifold of  ${\cal M}_n^0$ is now known, 
but at the present time this is the full extent of explicit results
for the exact metric. 
Using a point particle approximation the asymptotic metric on parts of the moduli space representing 
well-separated monopoles has been explicitly computed by Gibbons and Manton
\cite{GM}.

Imposing tetrahedral symmetry upon charge four monopoles gives,
after fixing the centre of mass and orientation, a totally geodesic
1-dimensional submanifold ${\cal N}\subset {\cal M}_4^0.$
The associated four monopole scattering has been investigated in detail
\cite{HSa} and the metric on ${\cal N}$ computed numerically \cite{S}.
In this letter we calculate this metric exactly, and in closed form,
in terms of complete elliptic integrals. As noted above, this is the first
 explicit calculation of the metric on any submanifold of ${\cal M}_n^0$,
except for the Atiyah-Hitchin submanifolds. These latter submanifolds
arise through the embedding of $n$ collinear monopoles and thus it is clear
that the intersection of the submanifold ${\cal N}$ with the above
Atiyah-Hitchin submanifold is empty. 

The approach taken is to construct the metric on the moduli space
of Nahm data, with the tangent vectors obtained by direct differentiation.
Comparisons with numerical results and the asymptotic metric are  then made.

\section{Tetrahedral charge four monopoles}
\news
The Nahm data
and spectral curve for the one-parameter family of 4-monopoles
described by ${\cal N}$ have been calculated \cite{HSa}. 
The spectral curve for this family is
\be
\eta^4+ i36a\omega^3\eta\zeta(\zeta^4-1)+
3\omega^4(\zeta^8+14\zeta^4+1)=0\label{sc}
\ee
where $a\in(-a_c,a_c)$, $a_c=3^{-5/4}\sqrt{2}$,
and $\omega$ is  the real half-period of the elliptic curve
\be
y^2=4(x^3-x+3a^2).
\ee
For the special point $a=0$ the monopole actually has 
cubic symmetry and was discovered by Hitchin, Manton
and Murray \cite{HMM}. In the limit
 as $a\rightarrow a_c$ we have $\omega\rightarrow \infty$.
Recalling that a single monopole with position
$(x_1,x_2,x_3)$ has spectral curve
$
\eta -(x_1+i\, x_2)+2 x_3\zeta +(x_1-i\, x_2)\zeta^2=0,
$
the product of the four spectral curves corresponding to monopoles
positioned on the vertices  
$\{ (-l,-l,-l)\frac{1}{\sqrt{3}},(-l,+l,+l)\frac{1}{\sqrt{3}},
   (+l,+l,-l)\frac{1}{\sqrt{3}}, (+l,-l,+l)\frac{1}{\sqrt{3}}\}$
of a large regular tetrahedron then has asymptotic spectral curve 
\be
\eta^4+ i\frac{16}{3^{3/2}}l^3\eta\zeta(\zeta^4-1)+
\frac{4}{9}l^4(\zeta^8+14\zeta^4+1)=0.\label{sca}
\ee

By comparing  (\ref{sc}) and (\ref{sca}) we see that 
we can make the identification 
\be
l=\Lambda a^{1/3}\omega, \hskip 1cm \mbox{where } 
\hskip 1cm \Lambda=3^{7/6}2^{-2/3}.
\ee
This $l\in R$ is a good global coordinate on ${\cal N}$: it
is zero when the monopoles coincide to form the cubic monopole 
and may be identified with the coordinates of the vertices of the tetrahedron
when $l$ is large. The task at hand is to compute the metric on  ${\cal N}$
in terms of $l$.
It is known that the transformation between the  monopole
moduli space metric and the metric on Nahm data is an
isometry \cite{Na} and so we may  construct the metric
on ${\cal N}$ by computing the  metric on the Nahm data.

The Nahm data for this family of monopoles has the form
\be
T_i(s)=x(s)X_i+y(s)Y_i+z(s)Z_i \hskip 20pt i=1,2,3
\label{tnd}
\ee
where $x,y,z$ are the real functions
\bea
&\left( x(s),y(s),z(s)\right)=
\left( 
\frac{\omega}{5}\left(-2\sqrt{\widetilde\wp(u)}+
\frac{1}{4\widetilde\wp(u)}\frac{d\widetilde\wp(u)}{du\ \ \ }\right),
\frac{\omega}{20}\left(\sqrt{\widetilde\wp(u)}
+\frac{1}{2\widetilde\wp(u)}\frac{d\widetilde\wp(u)}{du\ \ \ }\right),
\frac{a\omega}{2\widetilde\wp(u)}
\right).\nonumber
 \ \hskip -1cm \label{xyz}\\
\eea
Here $u=\omega s$ and $\widetilde\wp$ is the Weierstrass function satisfying
\be (\frac{d\widetilde\wp}{du})^2=4\widetilde\wp^3-4\widetilde\wp+12a^2.
\label{wfun}\ee
The tetrahedrally symmetric Nahm triplets $X_i,Y_i$ and $Z_i$ are constant
$4\times 4$ matrices; explicit expressions for these may be found in \cite{S}.
The spectral curve (\ref{sc}) is then related to this data by
setting
$\det(\eta\, I + (T_1+i T_2) -2 i\zeta T_3 + (T_1- iT_2) \zeta^2 )=0$.

Let $V_i=dT_i/dl$ be the tangent vector corresponding to the point with
Nahm data $T_i$. The metric on Nahm data is then given by
\be
g=-\Omega\int_0^2 \sum_{i=1}^3 \mbox{tr}(V_i^2)\ ds,
\label{mnd}
\ee
where tr denotes trace and $\Omega$ is a normalization constant.
In general a fourth Nahm matrix and its corresponding
tangent vector needs to be introduced to ensure orthogonality to gauge
orbits, but in this particular case the tetrahedral symmetry of the Nahm data
implies that this can be ignored and the resulting tangent vectors
are automatically orthogonal to the gauge orbits. (See \cite{S} for a discussion
and proof of this fact.) After substituting the explicit expressions
for the matrices $X_i,Y_i$ and $Z_i$ and performing the traces we
obtain
\be
g(l)=12\Omega\int_0^2\{
5(\frac{dx}{dl})^2+80(\frac{dy}{dl})^2+3(\frac{dz}{dl})^2\} ds.
\label{integral}
\ee
We calculate the quantity (\ref{integral}) in the next section.
Before turning to this however we need to make few remarks about
normalizations.

For a single monopole the length of the Higgs field
$\vert\Phi\vert$ has the asymptotic behaviour
\be
\vert\Phi\vert=v-\frac{g}{4\pi r}+O(e^{-8\pi vr/g})
\label{units}
\ee
where $v$ is the vacuum expectation value of the Higgs field
and $g$ is the magnetic charge. The monopole mass is the product
of these two constants, $m=vg$.
Performing the ADHMN construction for a single monopole we find
\be
\vert\Phi\vert=1-\frac{1}{2 r}+..
\ee
from which we can read off our units to be $m=g=2\pi$.

\section{Determining the metric}
\news
In this section we use several identities from the theory of elliptic
functions. These can be found in, for example, reference \cite{AS} 
and we follow their notation throughout.

As a first step towards evaluating (\ref{integral}) it is
helpful to disentangle the $l$ dependence in the arguments
of the functions  appearing in the integrand 
by exploiting the homogeneity properties of the $\wp$-function.
This enables us to express Nahm data in a more convenient form.  We have
\be
\widetilde\wp(\omega s)=\omega^{-2}\wp(s)
\ee
where $\wp(s)$ now satisfies the equation
 (throughout $'$ denotes differentiation with respect to $s$)
\be 
\wp^{\prime 2}=4\wp^3-g_2\wp-g_3
\label{peqn}
\ee
with the parameters
\bea
g_2&=&4(m^2-m+1)K^4/3\cr
g_3&=&4(m-2)(2m-1)(m+1)K^6/27\cr
\omega&=&[(m^2-m+1)/3]^{1/4}K.
\eea
Here $K$ denotes the complete elliptic integral with parameter $m$.
The functions in the Nahm data now take the simplified form
\begin{equation}
\left( x(s),y(s),z(s)\right)=
\left(\frac{1}{5}\left(-2\sqrt{\wp(
    s)}+\frac{1}{4}\frac{\wp^\prime( s)}{\wp(
    s)}\right),
\frac{1}{20}\left(\sqrt{\wp(
    s)}+\frac{1}{2}\frac{\wp^\prime( s)}{\wp( s)}\right),
\frac{\sqrt{-g_3}}{2\sqrt{12}\wp(s)}
\right).
\label{xyzsimp}
\end{equation}
All of the $l$ dependence resides in the parameters $g_2$ and
$g_3$.
In terms of the elliptic
parameter $m$ the geodesic coordinate $l$ may now be expressed as
\be
l^6=\frac{3^7}{2^4}a^2\omega^6=\frac{3^3}{2^4}\mmm K^6.
\label{ltom}
\ee
Because we have expressed all the parameters in terms of $m$ it is
convenient to perform  differentiations with 
respect to $m$ (which we denote by a dot) rather than $l$.
Accordingly we will evaluate
\be
\widehat g=g \dot l^2,
\label{newg}
\ee
where, upon using (\ref{ltom}), we find that
\be
\dot l=\frac{\sqrt{3}\left[\mmm E-K(1-m)(m^2+2m-2)\right]}
{m(1-m)\left[4\mmm\right]^{5/6}}.
\label{ldot}
\ee
Here $E$ is the complete elliptic integral of the second kind with
parameter $m$.

Upon substituting (\ref{xyzsimp}) into (\ref{integral}) and
noting (\ref{newg}) our task is now to evaluate
\be
\widehat g =\frac{3\Omega}{4}\int_0^2\{
\frac{4\dot \wp^2}{\wp}+(\frac{d}{ds}\frac{d}{dm}\log\wp)^2+
(\frac{d}{dm}\frac{\sqrt{-g_3}}{\wp})^2\} ds.
\label{ghat}
\ee
Using standard expressions for the partial derivatives of the $\wp$-function
with respect to the invariants $g_2$ and $g_3$ we obtain
\be
\dot\wp=\wp' H+F, \ \
\mbox{where}\ \
H= {a \zeta+b s}\ \mbox{and}\ \ 
F={\alpha \wp^2 +\beta \wp +\gamma}.
\label{pdot}
\ee
Here we have set
$$a= (3g_2  \dot g_3 -9 g_3 \dot g_2 /2)/\Delta , \ 
 b= (g_2 ^2 \dot g_2 /4 -9 g_3 \dot g_3/2)/\Delta $$
 \be
\alpha= (6 g_2  \dot g_3 -9 g_3 \dot g_2)/\Delta , \
\beta = (g_2 ^2 \dot g_2 /2 -9g_3 \dot g_3)/\Delta , \
\gamma= (3 g_2  g_3 \dot g_2 /2 -g_2^2 \dot g_3)/\Delta .
\ee
Of course, $\zeta$ is the Weierstrass $\zeta$-function satisfying
$
\zeta'=-\wp
$
and $\Delta$ is the discriminant $\Delta=g_2^3-27g_3^2$.

Substituting (\ref{pdot}) into (\ref{ghat}) and using (\ref{peqn})
together with its derivative we find
\be
\widehat g =\frac{3\Omega}{4}\int_0^2\ J\ ds
\ee
with the integrand $J$ given by
\newcommand{\ta}{((\alpha-a)\wp^2+b\wp-\gamma)}
\bea
&\wp^4J=H^2[20\wp^6-2g_2\wp^4-4g_3\wp^3+\frac{1}{4}g_2^2\wp^2+2g_2g_3\wp
+2g_3^2]\\
&+2H\wp'[\frac{1}{2}\dot g_3\wp+F(4\wp^3-g_3)+(2\wp^3+\frac{1}{2}g_2\wp+g_3)
\ta]\nonumber\\
&+F^2(4\wp^3-g_3)+(4\wp^3-g_2\wp-g_3)\ta^2+\dot g_3\wp(F-
\frac{1}{4}\dot g_3\wp/g_3).\nonumber
\eea
Now the potentially problematic terms are those involving $H$, since
this contains the $\zeta$-function. However, after some
calculation, we can express all the terms involving $H$ as total
derivatives ie.
\bea
J=\frac{d}{ds}\{
H^2[\frac{10}{3}\wp'+\frac{2}{3}g_3\wp'/\wp^3+\frac{1}{6}g_2\wp'/\wp^2]
+H[(6\alpha+\frac{4}{3}a)\wp^2+8(\beta-\frac{1}{3}b)\wp\nonumber\\
-\frac{2}{3}(g_2b-ag_3)/\wp
-\frac{1}{2}(\dot g_3-2\beta g_3-g_2\gamma+\frac{2}{3}bg_3)/\wp^2
+\frac{4}{3}\gamma g_3/\wp^3]\}\nonumber\\
+ P/\wp^4\hskip 10cm
\eea
where $P$ is a $7^{th}$ order polynomial in $\wp$.
Using the identity
$$
\frac{1}{\wp\sp{r}}=\frac{d}{ds}\left(\frac{1}{g_3\, (r-1)}
\frac{\wp'}{\wp\sp{r-1}}\right)+\frac{1}{g_3\,\wp\sp{r-3}}
\left(4-\frac{6}{r-1}\right)+
\frac{g_2}{g_3\,\wp\sp{r-1}}\left(\frac{1}{2 (r-1)}-1\right)
$$
valid for $r\ge2$ we then recursively remove the terms
$1/\wp\sp4$, $1/\wp\sp3$ and $1/\wp\sp2$ and we find the
difficult $1/\wp$-term vanishes. We are then left with a
polynomial of degree $3$ in $\wp$ which is readily integrated.
At this stage we have expressed the integrand  $J$ as the total derivative of a density ${\sl j}$, ie. $J=dj/ds$.
It now remains to evaluate the density at the limits of integration,
$s=0,2$.

As $s\rightarrow 0$ the required asymptotic limits are 
\be
\wp\sim \frac{1}{s^2}+\frac{1}{20}g_2s^2, \
\zeta\sim \frac{1}{s^2}-\frac{1}{60} g_2 s^3.
\ee
We find that the pole terms in the density cancel, which is a highly non-trivial
check on our calculation, and furthermore everything is proportional
to $s$, giving the result $j\vert_{s=0}=0$.

Finally, as $s\rightarrow 2$, we have
\be
\wp\sim \frac{1}{(s-2)^2}+\frac{1}{20}g_2(s-2)^2, \
\zeta\sim \frac{1}{s-2}+\frac{2}{3}K(3E+(m-2)K)-\frac{1}{60} g_2 (s-2)^3.
\ee
Again the pole terms in the density cancel, and we are left with a finite
value for $j\vert_{s=2}$ which is non-zero. Using (\ref{newg}) and
(\ref{ldot}) and choosing the normalization constant $\Omega=\pi$ we
obtain the final result
\bea
&g(l)=
8\pi \left( \frac{f(m)}{2}\right)\sp{2/3} 
\frac{ \left\{ 2 f(m) (2-m) E\sp2 -2f(m)E\sp3 /K -2f(m) (1-m) E K+m\sp2
               (1-m)\sp2 K\sp2
       \right\}
     }{ \left[ f(m) E -(m^2+2m-2) (1-m)K\right]\sp2}
\nonumber\\
& \ \hskip -1cm
\label{final}
\eea
where we have set
\be
f(m)=(2-m)(1+m)(2m-1)
\label{finalf}
\ee
and we recall that 
\be
l=\frac{\sqrt{3}}{2^{2/3}}[\mmm]^{1/6}K.
\ee

\section{Analysis of the metric}
\news
In figure 1 we plot (solid curve) the metric (\ref{final}) as a function
of $l$ for $l\in[0,6]$. 

\noindent Using a point particle approximation, Gibbons and Manton \cite{GM}
have calculated the asymptotic metric 
on regions of ${\cal M}_n$ which describe well-separated monopoles.
For pure monopoles ie. with zero electric charge, the asymptotic metric
for $n$ monopoles with positions ${\bf x}_i$ is given by
$ds^2=g_{ij}\dot{ \bf x}_i\cdot\dot {\bf x}_j$,
\be
g_{jj}=m-\frac{g^2}{4\pi}\sum_{i\ne j}
\frac{1}{\vert {\bf x}_i-{\bf x}_j\vert},\ \ \ \ 
g_{ij}=\frac{g^2}{4\pi}\frac{1}{\vert {\bf x}_i-{\bf x}_j\vert}.
\label{gm}
\ee
Here $m$ and $g$ are the mass and magnetic charge of a single monopole.
As stated earlier, in the normalization we have chosen these values are
$m=g=2\pi$. 
Note that the numerical construction
of the metric in reference \cite{S} involves a renormalisation
of $l\mapsto 2l$ in the plot of the metric. From equation
(\ref{units}) it can easily be seen that the effect of this 
scaling of the space coordinates is to convert to 
the more standard normalization of 
$m=g=4\pi$. 

For four monopoles on the vertices of a regular tetrahedron, as
described earlier, the asymptotic metric (\ref{gm}) becomes
\be
g_{GM}=8\pi (1-\frac{\sqrt{6}}{2l}).
\label{asy}
\ee
In figure 1 we also plot this metric (dashed curve) for comparison
with the true metric.

 It is known that in the asymptotic region
of large $l$ the correction to the metric  (\ref{asy}) is exponentially small.
Using the exact metric (\ref{final}) and taking the asymptotic limit,
which corresponds to $m\rightarrow 1$, we can calculate the leading
order correction to (\ref{asy}).

Introducing $m_1=1-m$ and using the standard expansions
\begin{eqnarray}
E=E(m)&=&1+\frac{m_1}{4}\left( \ln\frac{16}{m_1} -1\right)+
          \frac{3 m_1\sp2}{32} \left( \ln\frac{16}{m_1} -\frac{13}{6}\right)
        +\ldots \\
K=K(m)&=&\frac{1}{2}\ln\frac{16}{m_1}+
         \frac{m_1}{4}\left(\frac{1}{2}\ln\frac{16}{m_1} -1\right)+
         \frac{9m_1\sp2}{64} \left(\frac{1}{2}\ln\frac{16}{m_1} -\frac{7}{6}
         \right)+\ldots
\end{eqnarray}
we find that (working to quadratic order in $m_1$),
\bea
E&=&1+\frac{1}{4}(2K-1)m_1+\frac{1}{64}(4K-5)m_1^2\\
K&=&\frac{l\sqrt{2}}{\sqrt{3}}(1+\frac{1}{4}m_1+\frac{15}{32}m_1^2).
\eea
Upon substituting these expressions into (\ref{final}) the  terms linear 
in $m_1$ cancel and, after replacing $m_1$ by its leading order approximation
\be
m_1\sim 16\ \mbox{exp}(-l\sqrt{8}/\sqrt{3}),
\ee
the leading order correction to (\ref{asy}) is found to be
\be
\widetilde g=8\pi (1-\frac{\sqrt{6}}{2l}-192\sqrt{6}\ l\ 
\mbox{exp}(-2l\sqrt{8}/\sqrt{3})).\label{higher}
\ee
The first two terms in this expansion are readily identified as
coming from the first two terms in the numerator of (\ref{final}).

It is perhaps of interest to compare this exponential correction with
that in the charge two case, where there is an exponential
correction to the Taub-NUT metric to obtain the asymptotic 
Atiyah-Hitchin metric.

In the normalization we are using the asymptotic Atiyah-Hitchin metric
is \cite{GMa}
\be
\widetilde g_2=4\pi(1-\frac{2}{r}+8re^{-r})
\ee
where $r$ is the distance of each monopole from the centre of mass.
To compare this with the charge four case we
write $l=r\sqrt{6}/4$, after which (\ref{higher}) becomes
\be
\widetilde g/2=4\pi (1-\frac{2}{r}-288re^{-2r}).
\ee
We see that the exponential correction is a higher order effect in
the charge four case, which stems from the vanishing of the linear
terms in $m_1$. Presumably this arises since we are considering a
configuration which is particularly symmetric, leading to a cancellation
between the naive collection of two monopole pairs. Note also that
the correction to the Taub-NUT metric in the charge four case has
opposite sign to that in the charge two case.

Finally, in figure 1 we also plot some numerical values for the metric
(diamonds) computed using the algorithm introduced in \cite{S}.
This demonstrates that the numerical scheme is accurate, in fact the
computed values are all within $\frac{1}{2}\%$ of the true values,
and can be reliably used to compute the metric in other less tractable
cases.

\section{Conclusion}
\news
By direct calculation, and the use of explicit Nahm data, 
we have computed the metric on the moduli space of tetrahedrally
symmetric charge four BPS monopoles. An important ingredient was
the ability to write a combination of tangent vectors to Nahm data
as a total derivative. Note that Hitchin has remarked that a similar
situation may exist in general \cite{H}, with the charge $n$ metric being
determined by boundary values of a Riemann theta function on 
a surface of genus $(n-1)^2$. The difficulty in this formulation appears
to be the implementation of suitable boundary conditions on the 
general theta function solution of Nahm's equation. For the 
situation considered in this letter the tetrahedral symmetry implies
that it is the quotient surface under the tetrahedral group which is
relevant, and this has genus 1, thus allowing an explicit computation
in terms of elliptic functions and integrals.

\newpage
\begin{figure}[ht]
\begin{center}
\leavevmode
\vskip -4cm
{\epsfxsize=12cm \epsffile{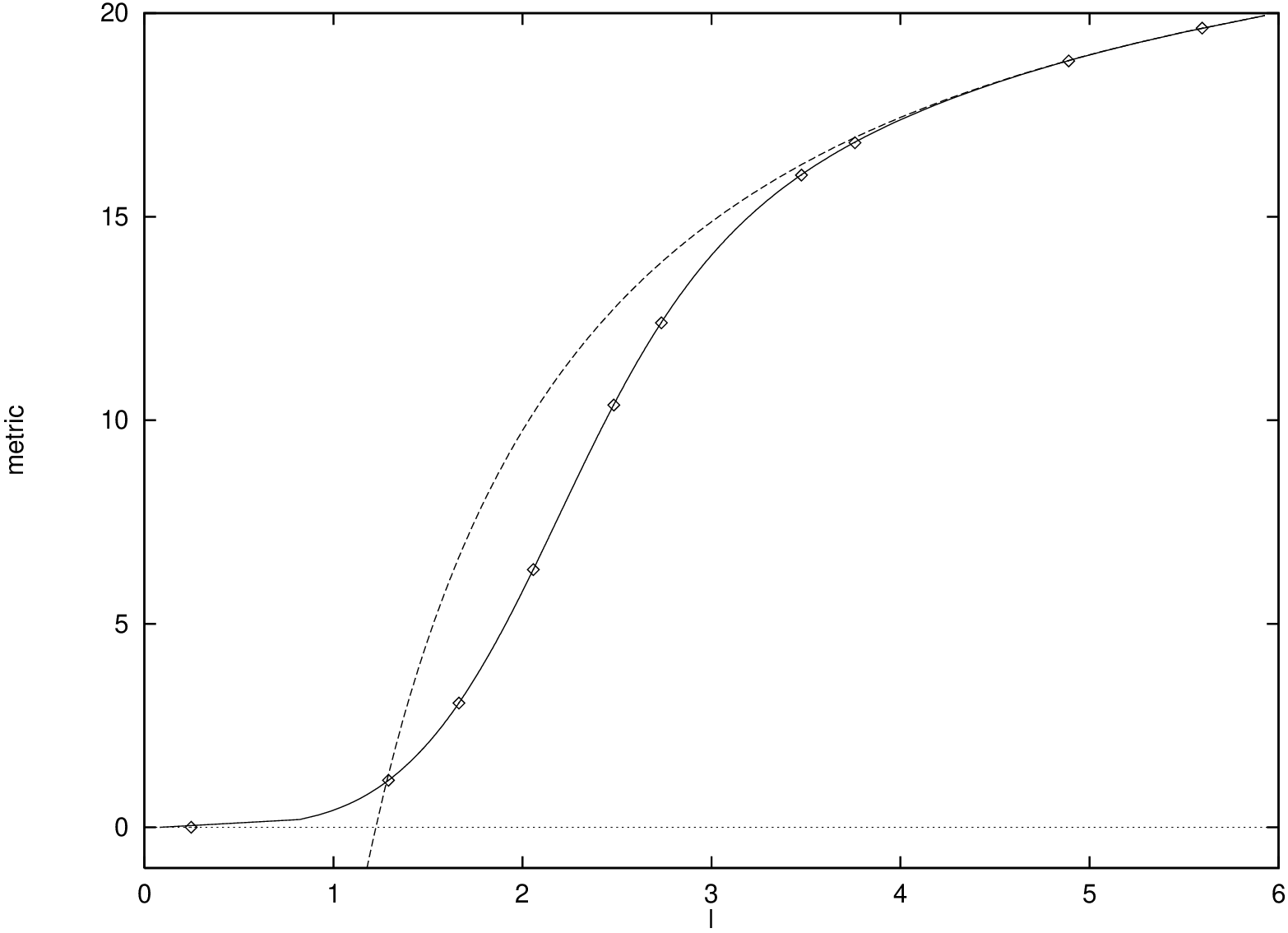}}
\vskip 2cm
\caption{The exact metric (solid curve), asymptotic metric
(dashed curve) and numerical results (diamonds).}
\end{center}
\end{figure}

\end{document}